\documentclass[10pt,amssymb]{revtex4}
\usepackage{graphicx}
\newcommand{\ket}[1]{|#1\rangle}
\newcommand{\bra}[1]{\langle#1|}

\newcommand{\mc}{\mathrm}

\begin{document}

\title{Work extraction from tripartite entanglement}
\author{Vincent Vigui\'{e}$^{1,2}$, Koji Maruyama$^{1,3}$, and Vlatko Vedral$^{4,5}$}
 \affiliation{$^1$QOLS, Blackett Laboratory, Imperial College London,
   London SW7 2BW, United Kingdom\\
$^2$D\'{e}partement de Physique, Ecole Normale Sup\'{e}rieure de
Lyon, 69364 Lyon, Cedex 07, France\\
$^3$Laboratoire d'Information Quantique and QUIC, CP 165/59, Universit\'{e} Libre de Bruxelles, Avenue F.D. Roosevelt 50, 1050 Bruxelles, Belgium\\
$^4$The School of Physics and Astronomy, University of Leeds, Leeds LS2 9JT, UK\\
$^5$Institut f\"{u}r Experimentalphysik, Universit\"{a}t Wien, Boltzmanngasse 5
1090 Wien, Austria}
\date{\today}

\begin{abstract}
The work extractable from correlated bipartite quantum systems can be used to distinguish
entanglement from classical correlation. A natural question is now whether it can be generalized
to multipartite  systems. In this paper, we devise a protocol to distinguish the GHZ, the W, and
separable states in terms of the thermodynamically extractable work under local operations and
classical communication, and compare the results with those obtained from Mermin's inequalities.
\end{abstract}
\maketitle

\section{Introduction}
Finding an efficient separability criterion for multipartite quantum systems and
characterizing entanglement have been an important problem in the field of quantum
information theory. Since Bell formulated the statistical irregularities that can be
seen in the correlations between measurement outcomes from two distant, but entangled,
systems \cite{bell64}, various criteria and related notions have been discovered in the
context of non-local correlations in quantum systems
\cite{horodecki96,horodecki99,nielsen01,terhal00}. Recently, it has been shown that it
is possible to devise a protocol to extract more work via entangled systems from a heat
bath thermodynamically than can be done from any separable state \cite{mmv}. This is of
interest because a physically useful quantity, locally extractable work, can be employed
to test the existence of entanglement in bipartite quantum systems, leading to the idea
of thermodynamical separability criteria.

In this paper, we discuss if the thermodynamical separability criteria can be generalized
to multipartite systems, particularly tripartite ones, as even the simplest transition
from bi- to tripartite systems makes our problems much harder. It has been known that,
unlike bipartite systems, there are two non-equivalent classes of entanglement, i.e. the
GHZ and the W states, when three quantum subsystems have non-local correlations
\cite{dur00}. Suppose that we are given an ensemble of tripartite systems, which is in
either the separable or the GHZ or the W states, where the GHZ state \cite{ghz89} can be
in general written as
\begin{equation}\label{ghz}
\ket{\mc{GHZ}}=\frac{1}{\sqrt{2}}(\ket{000}+\ket{111}),
\end{equation}
and the W state \cite{dur00},
\begin{equation}\label{w}
\ket{\mc{W}}=\frac{1}{\sqrt{3}}(\ket{001}+\ket{010}+\ket{100}),
\end{equation}
where $\ket{0}$ and $\ket{1}$ are two orthogonal states of a two-level system, e.g.
eigenstates of a Pauli spin operator $\sigma_z$. Our task here is to distinguish each of
these two ensembles from separable states in terms of extractable work. By extractable
work, we mean work that can be extracted thermodynamically from local heat baths under
local operations and classical communication (LOCC).

\section{The protocol for bipartite correlations}\label{bipartite}
Before discussing the case of tripartite entanglement, we here sketch the protocol shown
in \cite{mmv} for the work-extraction from bipartite quantum systems. If we have an
ensemble of two-level systems, either classical or quantum, we can extract work of amount
of $kT\ln2[1-H(X)]$ from a heat reservoir thermodynamically
\cite{vonneumann,bennett82,oppenheim02}, where $k, T$, and $H(X)$ are the Boltzmann
constant, the temperature of the heat bath, and $H(X)$ is the Shannon entropy of a
binary random variable $X$, respectively. The variable $X$ corresponds to the outcome of
measurement on the system and $H(X)$ can be written as $H(X)=-p_0\log_2 p_0-p_1\log_2
p_1$, where $p_0$ and $p_1$ are the probabilities for the two outcomes. We will set
$kT\ln2=1$ for simplicity hereafter and call the unit of work ``bit". We will also write
"extracting work from quantum systems" for short, instead of "extracting work
thermodynamically from a heat bath via quantum systems".

The extractable work from correlated pairs in \cite{mmv} is a simple generalization of
the above case. Suppose that two distant parties, Alice and Bob, have an ensemble of
identically prepared pairs of quantum bits (\textit{qubits}), which is described by a
density operator $\rho$. First, we define the extractable work $\xi_\rho(A(\theta),
B(\theta^\prime))$ when Alice and Bob chose $\theta$ and $\theta^\prime$ as the
directions of their (projective) measurement. After dividing the shared ensemble into
groups of two pairs, Alice measures one of the two qubits in a group with the projector
she chose and informs Bob of the outcome (See Fig. \ref{protocol}). Bob performs the
same on his qubit of the other pair in the group. As a result of collective
manipulations, they can extract $\xi_\rho(A(\theta), B(\theta^\prime))=1-1/2\cdot
[H(A(\theta)|B(\theta^\prime))-H(B(\theta^\prime)|A(\theta))]$ bits of work per pair at
maximum.

\begin{figure}
 \begin{center}
  \includegraphics[scale=0.3]{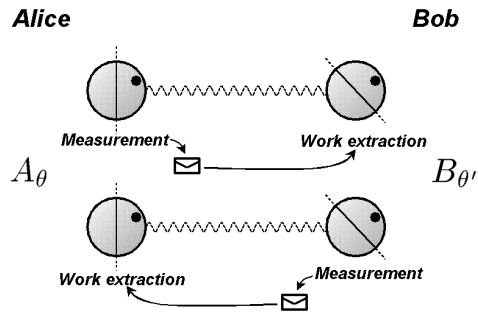}
  \caption{Schematic view of the protocol to extract work from correlated
pairs. Dividing the whole ensemble into groups of two pairs, Alice and Bob use $A(\theta)$ and
$B(\theta^\prime)$ for their measurement and work extraction. For a half of this ensemble, Alice
measures her state with $A(\theta)$ and Bob maximizes the extractable work from his side along the
direction of $\theta^\prime$ by using Alice's measurement results. For the other half, they
exchange their roles.}
 \label{protocol}
 \end{center}
\end{figure}

Second, we consider a quantity $\Xi(\rho)$, which is an average work extractable when we
set $\theta=\theta^\prime$ and vary $\theta$ continuously over a great circle on the
Bloch sphere:
\begin{equation}\label{var_basis}
\Xi(\rho):=\frac{1}{2\pi}\int_0^{2\pi}\xi_\rho(A(\theta), B(\theta))d\theta,
\end{equation}
The great circle is the one that maximizes the integral. The integral in Eq.
(\ref{var_basis}) can be taken over the whole Bloch sphere, nevertheless, it does not
change the essential part of our discussion here.

In \cite{mmv}, it was shown that an inequality
\begin{equation}\label{theineq}
\Xi(\rho)\le\Xi(\ket{00})=0.4427
\end{equation}
is a necessary condition for a two-dimensional bipartite state $\rho$ to be separable,
that is of the form, $\rho=\sum_i p_i \rho_i^A\otimes\rho_i^B$. The state $\ket{00}$ in
the right-hand side (RHS) can be any pure product state $\ket{\psi\psi'}$.

If we integrate $\xi_\rho$ over the whole surface of the Bloch sphere, then the condition
(\ref{theineq}) becomes
\begin{equation}\label{theineqBS}
\Xi_{BS}(\rho)\le\Xi(\ket{00})=0.2787,
\end{equation}
where the subscript $BS$ stands for the Bloch sphere. This condition, Eq.
(\ref{theineqBS}), has been shown to be more effective than the standard
Bell-Clauser-Horne-Shimony-Holt (Bell-CHSH) inequalities \cite{chsh} in detecting the
inseparability of the Werner state \cite{werner89}, which is a state in the form of
$\rho_W=p\ket{\Psi^-}\bra{\Psi^-}+(1-p)/4\cdot I$.

\section{Protocol for tripartite systems}\label{sec_pro3}
Let us now consider the case of tripartite quantum systems. By generalizing the protocol
in the previous section in a simple manner, we can have a necessary condition for
separability, which is of the same form as Eq. (\ref{theineq}), straightforwardly. That
is, each one of three parties receives the outcomes of measurements from the other two
and we take conditional entropies such as $H(A(\theta)|B(\theta), C(\theta))$ instead of
$H(A(\theta)|B(\theta))$ in the definition of $\xi$ and $\Xi$. Then, the inequality
(\ref{theineq}) with $\Xi(\ket{000})$ on the left-hand side (and a different numerical
value) holds for all separable states of the form of $\rho=\sum_i p_i
\rho_i^A\otimes\rho_i^B\otimes\rho_i^C$. The proof is essentially the same as that in
\cite{mmv}. Thus any violation of the inequality implies the existence of entanglement
between at least two subsystems. Obviously, this extension can be applied to
correlations between larger number of subsystems.

However, such an inequality tells little about the properties of multipartite
entanglement. Thus, we would like to find a different protocol and focus on
distinguishing the two inequivalent classes of tripartite entanglement: in this paper,
we will devise a way to distinguish the GHZ state, the W state, and separable states, in
terms of the extractable work.

With three parties (Alice, Bob, and Charlie), we consider a protocol in which one
extracts work along the direction of the $u$-axis after receiving information on the
outcomes of the other two's measurements along the $z$- and $u$- axes, respectively.
Namely, for a subensemble where Alice and Bob measure their qubits and Charlie extracts
work from his, the (average) amount of work obtainable is
\begin{equation}\label{tri_work01}
w_{\vec{z},\vec{u}}(\rho)=1-H(C(\vec{u})|A(\vec{z}),B(\vec{u})).
\end{equation}
In this equation, we denote the direction of axis by a unit vector, such as $\vec{z}$.
We have chosen two directions, $\vec{z}$ and $\vec{u}$, for two measurements in
anticipation that the difference between the GHZ and the W states will be seen by
varying the direction of $\vec{u}$, while keeping $\vec{z}$ fixed. We do not lose
generality by choosing the direction of the work-extraction to be the same as one of the
measurement axes, that is $\vec{u}$ in Eq. (\ref{tri_work01}), as it is the right choice
in detecting singlet-type bipartite entanglement. Also, as we focus on the GHZ and the W
states, which are symmetric in terms of $A, B,$ and $C$, we will assume throughout the
paper that it is Charlie who extracts work after Alice and Bob make measurements on
their subsystems. Otherwise, we need to permute the roles of each party and take an
average.

\begin{figure}
 \begin{center}
  \includegraphics[scale=0.5]{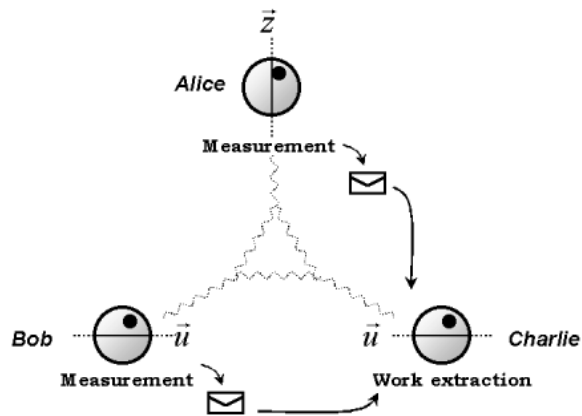}
  \caption{Work extraction from a tripartite system. Alice and Bob measure their particle along the direction of $\vec{z}$ and $\vec{u}$, respectively, where Bob varies the direction of $\vec{u}$, choosing one of the three orthogonal directions $\{\vec{x},\vec{y},\vec{z}\}$ at each round. Charlie extracts work along $\vec{u}$ after receiving measurement results from Alice and Bob. They exchange their roles cyclically after three rounds, however, we do not consider the exchange of roles as the states of our interest here are symmetric with respect to three parties.}
 \label{protocol}
 \end{center}
\end{figure}

We now show that a quantity below, $W(\rho)$, achieves our task. We assume that the
direction $\vec{z}$ is given and fixed for everyone of three parties throughout our
discussion. Instead of varying $\vec{u}$ continuously, we average the extractable work
over three orthogonal directions in the Bloch sphere, $(\vec{x}, \vec{y}, \vec{z})$, for
$\vec{u}$. This is mainly because of our limited resource for numerical computation, but
it does not change our main task at all. With only the direction of $\vec{z}$ fixed,
those of $\vec{x}$ and $\vec{y}$ cannot be determined. We use $\phi$ to specify the
angle between $\vec{x}$ and a certain predetermined direction in the space. Then, the
averaged work can be written as
\begin{equation}\label{tri_work02}
W_\phi(\rho)=\frac{1}{3}\sum_{\vec{u}\in \lbrace \vec{x}, \vec{y}, \vec{z}\rbrace} w_{\vec{z},\vec{u}}(\rho),
\end{equation}
and we take the maximum value of $W_\phi(\rho)$ over $\phi$ to
remove the $\phi$-dependence as
\begin{equation}\label{tri_work03}
W(\rho)=\max_{\phi}W_\phi(\rho).
\end{equation}
We will analyse how we can use $W(\rho)$ to distinguish three classes of correlations in
tripartite systems.

\subsection{Product states}
With a product state, $\ket{\psi^A\psi^B\psi^C}$, the extractable
work depends only on the subsystem from which we extract work,
regardless of the measurements on the other two. Thus,
$W(\ket{\psi^A\psi^B\psi^C})$ will be determined solely by the
geometric relation between the state of the system and the choice
of axes. $W_\phi(\ket{\psi^A\psi^B\psi^C})$ takes its maximum
value of $1/3$ when the Bloch vector representing the state
coincides with one of the axes, i.e.
$W(\ket{\psi^A\psi^B\psi^C})\le 1/3$ for a given $\vec{z}$.

\subsection{Separable states}
We now show that $1/3$ is indeed an upper bound for the extractable work $W$ from any
separable state. That is, an inequality,
\begin{equation}\label{tri_criterion}
W(\rho)\le \frac{1}{3},
\end{equation}
is a necessary condition for a tripartite state $\rho$ to be
separable. This condition is very similar to the thermodynamical
separability criterion for bipartite correlations \cite{mmv},
however, $W(\rho)$ is an average of work over only three
directions as in Eq. (\ref{tri_work02}), while it was over a
great circle on the Bloch sphere (or the whole surface of the
sphere) in \cite{mmv}. This results in a difference in the
efficiency of detecting weakly entangled states: The condition
(\ref{tri_criterion}) is less efficient than the criterion in
\cite{mmv}, in this sense.

\textit{Proof.} The conditional entropy in Eq. (\ref{tri_work01}) can be written as
\begin{equation}\label{tri_proof01}
H(C(\vec{u})|A(\vec{z}),B(\vec{u}))=\sum_{i,j=\{0,1\}}
p(A_{\vec{z}}^i,B_{\vec{u}}^j)H(C(\vec{u})|A_{\vec{z}}^i,B_{\vec{u}}^j),
\end{equation}
where $A_{\vec{z}}^i \; (B_{\vec{u}}^j)$ means that Alice (Bob) obtained the outcome
$i\,(j)$ by the measurement along $\vec{z}\;(\vec{u})$ and
$p(A_{\vec{z}}^i,B_{\vec{u}}^j)$ is the probability of a joint event of $A_{\vec{z}}^i$
and $B_{\vec{u}}^j$. Omitting the directions of axes for simplicity, we can write Eq.
(\ref{tri_proof01}) as
\begin{equation}\label{tri_proof02}
H(C|A,B)=\sum_{i,j}p_{ij}H(C|A^i,B^j).
\end{equation}
If the state $\rho$ is separable, i.e. $\rho=\sum_n
p_n\rho_n^A\otimes\rho_n^B\otimes\rho_n^C$, then the density operator for Charlie after
Alice and Bob obtained $i$ and $j$ becomes
\begin{equation}\label{tri_proof03}
\rho_{ij}^C=\frac{1}{p_{ij}}\sum_n p_n \mc{Tr}(P_z^i \rho_n^A)\mc{Tr}(P_u^j \rho_n^B)\rho_n^C,
\end{equation}
where $P_z^i\;(P_u^j)$ is a projection operator for outcome $i\;(j)$ along the direction
of $\vec{z}\;(\vec{u})$ and $p_{ij}=\sum_n p_n \mc{Tr}(P_z^i \rho_n^A)\mc{Tr}(P_u^j
\rho_n^B)$. Therefore,
\begin{eqnarray}\label{tri_proof04}
H(C|A,B) &=& \sum_{i,j} H\left( \frac{1}{p_{ij}}\sum_n p_n \mc{Tr}(P_z^i \rho_n^A)\mc{Tr}(P_u^j \rho_n^B)\mc{Tr}(P_u^0\rho_n^C)\right) \nonumber \\
&\ge & \sum_n \sum_{i,j}p_n\mc{Tr}(P_z^i \rho_n^A)\mc{Tr}(P_u^j \rho_n^B) H(\rho_n^C) \nonumber \\
&=& \sum_n p_n H(\rho_n^C),
\end{eqnarray}
because of the concavity of the Shannon entropy. If we were sure that we had a
tripartite product state, $\rho_k^A\otimes\rho_k^B\otimes\rho_k^C$, the conditional
entropy would be
\begin{eqnarray}\label{tri_proof05}
H(C^k|A^k,B^k) &=& H\left(\frac{1}{p_{ij}^k}\mc{Tr}(P_z^i \rho_k^A)\mc{Tr}(P_u^j \rho_k^B)\rho_k^C\right) \nonumber \\
&=& H\left(\rho_k^C\right),
\end{eqnarray}
where $p_{ij}^k=\mc{Tr}(P_z^i \rho_k^A)\mc{Tr}(P_u^j \rho_k^B)$.

Since $W(\rho_k)=\max[1-(1/3)\sum_{\vec{u}}H(C^k|A^k,B^k)]\le 1/3$ for all separable
states $\rho_k$, combining Eqs. (\ref{tri_proof04}) and (\ref{tri_proof05}), we have
\begin{eqnarray}\label{tri_proof06}
W(\rho) &=& \max\left[1-\frac{1}{3}\sum_{\vec{u}} H(C|A,B)\right] \nonumber \\
&\le & \max\left[1-\frac{1}{3}\sum_n \sum_{\vec{u}} p_n H(C^n|A^n,B^n)\right] \nonumber \\
&\le & \frac{1}{3},
\end{eqnarray}
thus Eq. (\ref{tri_criterion}). As $W(\rho)$ still has a dependence on the direction
of $\vec{z}$, even a highly entangled state may not violate the inequality
(\ref{tri_criterion}), however, once it is violated it is surely a manifestation of
entanglement.

\subsection{GHZ and W states}
Let us discuss how we can distinguish the GHZ, and the W states, using the above
criterion. To this end, we make use of the remaining variable, i.e. the direction of
$\vec{z}$.

The extractable work from a GHZ state can reach 1 bit, which is the maximum possible
violation of the inequality (\ref{tri_criterion}). This maximum is attained when we
choose $\vec{z}$ to be perpendicular to $\ket{0}$ in the Bloch sphere (such as
$\ket{+}=(\ket{0}+\ket{1})/\sqrt{2}$), the state after Alice's measurement still has
maximum (bipartite) entanglement. Because of the perfect correlation between $B$ and $C$,
Charlie can extract maximum work, 1 bit per one set of three subsystems.

We have obtained the value of the minimum extractable work from the GHZ as 0.1619 bits
numerically. This occurs when $\theta=1.5560$ and $\phi=0.5600$, for example, where
$\theta$ and $\phi$ are the azimuthal and the longitudinal angles in the Bloch sphere for
the direction of Alice's measurement basis $\vec{z}$. That is, the direction denoted by
$\theta=0$ is that of $\ket{0}$ and $\phi$ is a relative phase between $\ket{0}$ and
$\ket{1}$ in their superpositions.


The maximum work extractable from $\ket{\mc{W}}$ is $7/9$, which is attained when the
$\vec{z}$-direction is parallel to $\ket{0}$. If the outcome of Alice's measurement is 0,
which occurs with probability $2/3$, the other two systems, $B$ and $C$, will still have
a perfect correlation along the three orthogonal directions, being in the state
$(1/\sqrt{2})(\ket{01}+\ket{10})$ and thus Charlie can get 1 bit of work. If Alice's
outcome is 1, then Charlie can obtain only $1/3$ bits on average as the remaining two
systems are in a product state $\ket{11}$. Therefore, the average extractable work will
be $(2/3)\cdot 1+(1/3)\cdot (1/3)=7/9$ bits. The minimum work from the W state was
obtained numerically as 0.1696 bits. This is achieved when $\theta=0.7169$ and
$\phi=\pi/4$ for the direction of $\vec{z}$.

Now we can distinguish the GHZ and the W states, looking at the maximum and the minimum
values of $W$ after varying the direction of $\vec{z}$. The maximum and minimum
extractable work from the GHZ and the W states have the following relationships:
\begin{eqnarray}\label{ghzandw}
\frac{1}{3}< \max_{\vec{z}} W(\ket{\mc{W}})< \max_{\vec{z}} W(\ket{\mc{GHZ}}), \nonumber \\
\mbox{and}\hspace{2mm} \min_{\vec{z}} W(\ket{\mc{GHZ}})< \min_{\vec{z}}
W(\ket{\mc{W}}),
\end{eqnarray}
although the difference between the two minima is rather small. Therefore, when the state
of a given ensemble is one of the three possibilities, the GHZ, the W, and separable
states, as we have assumed above, it is possible to specify the state by examining the
range of extractable work. If $W(\rho)\le 1/3$ always holds regardless of the direction
of $\vec{z}$, then it is in a separable state due to the above proposition.

Suppose that entanglement exists only in two subsystems out of three, such as
$\sigma=\sigma^{AB}\otimes\sigma^C$, where $\sigma^{AB}=\ket{\Psi^-}\bra{\Psi^-}$ is a
maximally entangled state. If we wish to discriminate such bipartite entanglement in
tripartite systems, we need to examine the extractable work from each party, instead of
simply computing the average $W$. In this example of state $\sigma$, Alice and Bob can
obtain the maximum work (1 bit), regardless of the outcome of Charlie's measurement,
because of the perfect correlation between $A$ and $B$. However, no information from
Alice and Bob can be useful to maximize the work Charlie can extract. Thus, he can have
only $1/3$ bits of work at maximum after averaging over three directions. If one of the
subsystems is disentangled from the other two, the extractable work from this site cannot
exceed $1/3$, which is precisely what has been proved in the preceding subsection. By
comparing the extractable work from each site, we can distinguish tripartite correlation
with bipartite entanglement.

As we have mentioned above, the thermodynamical separability criterion for bipartite
systems is able to detect more inseparability of the Werner state of two qubits than the
Bell-CHSH inequalities. Let us consider a Werner-type state, in analogy of the bipartite
Werner state, with tripartite systems \cite{murao98}: We define the GHZ-Werner-type
state as
\begin{equation}\label{ghz_werner}
\rho_{W}^{\mc{GHZ}}=p\ket{\mc{GHZ}}\bra{\mc{GHZ}}+\frac{1-p}{8}\cdot I,
\end{equation}
and the W-Werner-type state as
\begin{equation}\label{w_werner}
\rho_{W}^{\mc{W}}=p\ket{\mc{W}}\bra{\mc{W}}+\frac{1-p}{8}\cdot I.
\end{equation}
These states are also referred to as isotropic states \cite{rains99}. The inequality
(\ref{tri_criterion}) turns out to be violated when $p>0.6521$ with the GHZ-Werner-type
state, and when $p>0.6981$ with the W-Werner-type state. Are these criteria better than
other schemes of multipartite-entanglement-detection? In \cite{murao98}, the
GHZ-Werner-type state with $p\ge 0.3226$ has been shown to be distillable, thus
entangled, where $p=(8f-1)/7$ with $f=\bra{\mc{GHZ}}\rho_{W}^{\mc{GHZ}}\ket{\mc{GHZ}}$ in
\cite{murao98}, while we have no corresponding data for the W-Werner-type state.

Let us take Mermin's inequality as the multipartite version of Bell-CHSH inequalities
\cite{mermin90}. Mermin's inequality for a tripartite system can be written as
\begin{equation}\label{mermin_ineq01}
\langle\mathcal{B}_3\rangle=\mc{Tr}(\rho\mathcal{B}_3)\le 2,
\end{equation}
where the operator $\mathcal{B}_3$ is defined as (by omitting the $\otimes$ sign)
\begin{equation}\label{mermin_ineq02}
\mathcal{B}_3=(\sigma^A_1\sigma^B_{2^\prime}+\sigma^A_{1^\prime}\sigma^B_2)\sigma^C_3
+(\sigma^A_1\sigma^B_2-\sigma^A_{1^\prime}\sigma^B_{2^\prime})\sigma^C_{3^\prime}.
\end{equation}
In Eq. (\ref{mermin_ineq02}), $\sigma_i^x=\vec{a_i}\cdot\vec{\sigma}^x$ is a measurement
operator at site $x\in\{A,B,C\}$ and $\vec{a_i}$ is a unit vector representing the
direction of the measurement.

The GHZ state, Eq. (\ref{ghz}), can violate the inequality (\ref{mermin_ineq01})
maximally as $\langle\mathcal{B}_3\rangle_{\mc{GHZ}}=4$. This means that the
GHZ-Werner-type state, Eq. (\ref{ghz_werner}), can violate Eq. (\ref{mermin_ineq01})
when $p>1/2$ since
\begin{equation}\label{ghz_werner_ineq}
\langle\mathcal{B}_3\rangle_{\mc{GHZ-Werner}}=\mc{Tr}(\rho_W^{\mc{GHZ}}\mathcal{B}_3)
=p\langle\mathcal{B}_3\rangle_{\mc{GHZ}}.
\end{equation}
Similarly, the W-Werner-type state, Eq. (\ref{w_werner}), can violate the inequality when
$p>0.6566$ as the maximum value of $\langle\mathcal{B}_3\rangle_{\mc{W-Werner}}$ is
3.046 \cite{cabello02}. Therefore, the condition (\ref{tri_criterion}) is less effective
in detecting the entanglement in both the GHZ-Werner-type and the W-Werner-type states:
There are a class of weakly entangled states that violate Mermin's inequality, but do
not violate Eq. (\ref{tri_criterion}).

\begin{table}
\caption{The comparison of the thermodynamical separability criterion for tripartite systems, Eq. (\ref{tri_criterion}), and Mermin's inequality, Eq. (\ref{mermin_ineq01}), in detecting the inseparability of the Werner-type states. This table shows the minimum values of $p$ to violate the inequalities, i.e. the smaller value, the more detection of inseparability.}\label{listofps}
\begin{center}
\begin{tabular}{c||c|c}\hline
  & Therm. sep. criterion (\ref{tri_criterion}) & Mermin's ineq. (\ref{mermin_ineq01}) \\ \hline
GHZ-Werner-type state & $p>0.6521$ & $p>0.5$\\
W-Werner-type state & $p>0.6981$ & $p>0.6566$\\
\hline
\end{tabular}
\end{center}
\end{table}

The situation does not change much even if we average the extractable work by varying the
$\vec{u}$ over the whole Bloch sphere. The inequality (\ref{tri_criterion}) can be
violated by the GHZ-Werner-type state when $p>0.8392$ and by the W-Werner-type state when
$p>0.9057$, after choosing $\vec{z}$ optimally. The reason why Eq. (\ref{tri_criterion})
is not as effective as Mermin's inequality is yet unclear, while a similar
thermodynamical criterion is more effective in the case of bipartite entanglement as
shown in \cite{mmv}.

\section{Summary}
We have generalized the thermodynamical separability criterion to tripartite quantum systems and
shown that it can be used to distinguish two different classes of tripartite entanglement, the
GHZ and the W, in terms of thermodynamically extractable work under LOCC. We have also found that
the criterion for tripartite systems is less effective in detecting the Werner-type entanglement
than that for bipartite systems. Although it is not perfectly clear if the Werner-type states
above are the proper counterpart of the bipartite Werner state we should compare with, it appears
that the separability criterion we have obtained here is not as efficient as other criteria, such
as Mermin's inequality, in detecting weak entanglement. Nevertheless, it is capable of
distinguishing the GHZ, the W, and separable states with physical quantity and this may lead to
an interesting physical or information processing process, in which the W state is more useful
than the GHZ state, unlike most of the known processes.

\section*{Acknowledgments}
We gratefully acknowledge discussions with \v{C}aslav Brukner. Vigui\'{e} is grateful to
Peter Holdsworth for arranging his stay in London. Maruyama is a Boursier de l'ULB, and
is supported by the European Union through the project RESQ IST-2001-37559, the Action de
Recherche Concert\'{e}e de la Communaut\'{e} Fran\c{c}aise de Belgique, and the IUAP
program of the Belgian Federal Government under grant V-18. Vedral acknowledges funding
from EPSRC and the European Union.

\end{document}